\documentclass[a4paper,12pt]{article}
\UseRawInputEncoding
\usepackage[left=2.5cm,bottom=3cm,right=2.5cm,top=3cm]{geometry} 
\usepackage{overpic,youngtab}
\usepackage{subfigure}
\usepackage[latin,english]{babel}
\usepackage{amsmath}
\usepackage{amssymb}
\usepackage{epstopdf}
\usepackage{graphics,psfrag,rotating}
\usepackage{mathabx}
\usepackage{graphicx}
\usepackage{dcolumn}
\usepackage{float}
\usepackage{pdflscape}
\usepackage{array}
\usepackage{booktabs}
\usepackage{amscd} 
\usepackage{mathtools}
\usepackage{fancybox}
\usepackage{fix-cm}
\usepackage[colorlinks=true,citecolor=blue,,linktocpage=true,linkcolor=blue,urlcolor=black]{hyperref}
\usepackage{tikz} 
\usetikzlibrary{matrix} 
\usetikzlibrary{positioning} 
\usetikzlibrary{arrows}
\usepackage{titlesec}
\usepackage{abstract}

\def\be{\begin{equation}}
\def\ee{\end{equation}}
\def\bea{\begin{eqnarray}}
\def\eea{\end{eqnarray}}
\def\bpm{\begin{pmatrix}}
\def\epm{\end{pmatrix}}

\def\a{\alpha}
\def\b{\beta}

\def\G{\Gamma}

\def\d{\delta}
\def\m{\mu}
\def\n{\nu}
\def\t{\tau}

\def\l{\lambda}

\def\r{\rho}

\def\ona{\overline{\nabla}}
\def \oR{\overline{R}}

\def\bR{\bar{R}}

\def\bn{\bar{\nabla}}
\def\bR{\bar{R}}
\def\og{\overline{g}}

\def\s{\sigma}

\def\bi{\begin{itemize}}
	\def\ei{\end{itemize}}
\def\bg{\bar{g}}

\begin{document}
	
	\vspace*{-1cm}
\phantom{hep-ph/***} 
{\flushleft
	{{FTUAM-21-xx}}
	\hfill{{ IFT-UAM/CSIC-21-96}}}
\vskip 1.5cm
\begin{center}
	{\LARGE\bfseries  Generalized Kerr-Schild gauge.}\\[3mm]
	\vskip .3cm
	
\end{center}

\vskip 0.5  cm
\begin{center}
	{\large Enrique \'Alvarez and Jes\'us Anero.}
	\\
	\vskip .7cm
	{
		Departamento de F\'isica Te\'orica and Instituto de F\'{\i}sica Te\'orica, 
		IFT-UAM/CSIC,\\
		Universidad Aut\'onoma de Madrid, Cantoblanco, 28049, Madrid, Spain
		\vskip .1cm

		\vskip .5cm
		
		\begin{minipage}[l]{.9\textwidth}
			\begin{center} 
				\textit{E-mail:} 
				\tt{enrique.alvarez@uam.es} and 
				\tt{jesusanero@gmail.com}
			\end{center}
		\end{minipage}
	}
\end{center}
\thispagestyle{empty}

\begin{abstract}
	\noindent
	The Kerr-Schild gauge is generalized to the case that the vector generating the deformation is not null. Contrary to naive expectations, this vector generates a finite expansion for the curvature tensor. We prove a theorem
	on the conditions for the deformed metric being Ricci flat, namely that the deformation vector must be irrotational (then geodesic) in the background spacetime.
	
\end{abstract}

\newpage
\tableofcontents
\thispagestyle{empty}
\flushbottom

\newpage

\section{Introduction}
One of the main problems in General Relativity is the non-linearity of the field equations, which means that the linear combination of solutions is not necessarily a solution, even at the free level. Given two Ricci flat metrics
\be
R_{\a\b}[g^{(1)}_{\m\n}]=R_{\a\b}[g^{(2)}_{\m\n}]=0
\ee
the sum is not Ricci flat anymore
\be
R_{\a\b}[g^{(1)}_{\m\n}+g^{(2)}_{\m\n}]\neq 0
\ee
this non-linearity can be traced back  the nonlinearity of the connection
\be
\G^\l_{\a\b}\left[g^{(1)}_{\m\n}+g^{(2)}_{\m\n}\right]\neq\G^\l_{\a\b}\left[g^{(1)}_{\m\n}\right]+\G^\l_{\a\b}\left[g^{(2)}_{\m\n}\right]
\ee
it is true that even if the connection were additive, the Ricci tensor would not share this property; but it seems a necessary step in the good direction. The main technical problem here  is the non additivity of   the inverse of a sum of matrices
\be
(g^{(1)}_{\m\n}+g^{(2)}_{\m\n})^{-1}\neq (g^{(1)}_{\m\n})^{-1}+(g^{(2)}_{\m\n})^{-1}
\ee

Let us examine this fact in more detail trying to discover exceptions to it; that is,
situations where the expansion of the inverse matrix terminates. To be specific, when
doing perturbation theory around a background gravitational field $\bg_{\m\n}$ 
\be
g_{\m\n}=\bg_{\m\n}+\kappa  h_{\m\n}
\ee
the first step in the computation is to calculate the inverse metric, $g^{\m\n}$. This is only possible as an infinite power series, namely
\be g^{\m\n}=\bg^{\m\n}-\kappa h^{\m\n}+\kappa^2 h_{~\t}^\m h^{\t\n}+\mathcal{O}(h^2)\ee
let us note that indices are raised with the $\bg_{\m\n}$ metric, there is no closed formula in the general case. The same thing happens with the determinant of the metric.

A clever  way out is the Kerr-Schild gauge \cite{Alvarez, Gurses, Kerr, Xanthopoulos}
\be \label{KS}g^{\m\n}=\bg^{\m\n}-\kappa h^{\m\n}\ee
where the deformation obeys 
\bea
&&h_{\m\l}h^{\l\n}=0\nonumber\\
&&h^\l_\l=0
\eea
these equations imply that 
\be h_{\a\b}=l_\a l_\b\ee
with $l_\a$ a null vector, $l^2=0$. In this case the Ricci tensor can be written as
\be R_{\m\n}=\bR_{\m\n}+R_{\m\n}^{\text{\tiny{(1)}}}+R_{\m\n}^{\text{\tiny{(2)}}}+R_{\m\n}^{\text{\tiny{(3)}}}\ee
where
\bea
&&R_{\m\n}^{\text{\tiny{(1)}}}=\frac{\kappa}{2}\left[\bn_\l\bn_\m h_{\n}^{~\l}+\bn_\l\bn_\n h_{\m}^{~\l}-\bar{\Box}h_{\m\n}\right]\nonumber\\
&&R_{\m\n}^{\text{\tiny{(2)}}}=\frac{\kappa^2}{2}\Big[\bn_\s\left[h^{\s\r}\bn_\r h_{\m\n}\right]-\bn_\s h_{\m}^{~\r}\bn_\r h_{\n}^{~\s}+\bn^\r h_{\m}^{~\s}\bn_\r h_{\n\s}-\frac{1}{2}\bn_\m h_{\r\s}\bn_\n h^{\r\s}\Big]\nonumber\\
&&R_{\m\n}^{\text{\tiny{(3)}}}=-\frac{\kappa^3}{2}h^{\r\s}\bn_\r h_{\m\l}\bn_\s h_{\n}^{~\l}\eea
The first term is essentially the Fierz-Pauli \cite{Fierz} equation for traceless spin 2 particles in flat space.
Hereinafter we will use
\be FP_{\m\n}\equiv \frac{1}{2}\left[\bn_\l\bn_\m h_{\n}^{~\l}+\bn_\l\bn_\n h_{\m}^{~\l}-\bar{\Box}h_{\m\n}\right]\ee
in the references  it was proved that
\be\label{teorema}
\{ R_{\m\n}^{\text{\tiny{(1)}}}=0\} \Longrightarrow \{R_{\m\n}^{\text{\tiny{(2)}}}+R_{\m\n}^{\text{\tiny{(3)}}}=0\}
\ee
This means that when the deformation  $h_{\m\n}$ obeys the Fierz-Pauli equation \cite{Fierz} (please remember  that $h^\l_\l=0$) then 
\be R_{\m\n}^{\text{\tiny{(1)}}}=\kappa FP_{\m\n}=0\ee
implies
\be R_{\m\n}=\bR_{\m\n}\ee
In the next section, we will present a completely different alternative.
\section{The non null Kerr-Schild gauge}
In the present work we would like to  generalize a little bit this setup by  assuming a finite expansion  for the inverse metric  in terms of a parameter $\xi$
\be \label{metricg}g^{\m\n}=\bg^{\m\n}+\xi\kappa h^{\m\n}\ee
where $\xi$ is an arbitrary constant. In fact we will be able to get $g_{\m\l}g^{\n\l}=\d_\m^\n$ provided 
\bea \label{rh}
&&\kappa h_{\m\a}\bg^{\a\b}h_{\b\n}=-\left(1+\frac{1}{\xi}\right)h_{\m\n}\nonumber\\
&&\kappa\bg^{\a\b}h_{\a\b}=-\left(1+\frac{1}{\xi}\right)\eea
this idea was initially proposed in the paper \cite{Gurses:2018ckx}.

On the other hand, the determinant $g$, with the matrix determinant lemma \cite{Harville}, is related to the background determinant $\bg$
\be g=\bg(1+\kappa\bg^{\m\n}h_{\m\n})=-\frac{1}{\xi}\bg\ee
note, if $\xi>0$, we can change the signature of the metric.

It is easy to check that there is also a relationship between the determinant of the
deformation $h_{\m\n}$ and the determinant of the background metric  $\bg_{\m\n}$
\be \kappa^4 h^2=\left(1+\frac{1}{\xi}\right)^4\bg h\ee 
there are two solutions. Either $h=0$, or else
\be h=\frac{1}{\kappa^4}\left(1+\frac{1}{\xi}\right)^4\,\bg\ee
but this corresponds with a rescaling of the metric
\be h_{\m\n}=-\frac{1}{\kappa}\left(1+\frac{1}{\xi}\right)\bg_{\m\n}\ee
and the total metric is
\be g_{\m\n}=-\frac{1}{\xi}\bg_{\m\n}\ee

Forgetting about this trivial rescaling, we will concentrate on the case $h=0$. This
means that the matrix  $h$ has got rank $r \leq 3$. Therefore, the condition \eqref{rh} has three different solutions. Let us concentrate in the simplest, namely $r=1$, which corresponds to

\be
h_{\m\n}=\kappa A_\m A_\n
\ee
There is a three-dimensional subspace of zero modes, defined by the condition
\be
A_\m n_i^\m=0\quad i=1,2,3
\ee
the condition \eqref{rh}  is fulfilled as long as 
\be\label{c}
\bg^{\a\b}A_{\a}A_{\b}=A^\l A_\l=-\frac{1}{\kappa^2 }\left(1+\frac{1}{\xi}\right)
\ee
in conclusion the metric reads
\be g_{\m\n}=\bg_{\m\n}+\kappa^2 A_\m A_\n\ee
and the inverse of the metric
\be g^{\m\n}=\bg^{\m\n}+\xi\kappa^2 A^\m A^\n\ee
in particular the determinant
\be g=\left(1+\kappa^2 A^\l A_\l\right)\bg\ee
This shows that for $\kappa^2A^2+1 > 0$ (that is, for timelike or null vector fields, and even for spacelike ones with $\kappa^2 A^2 > -1$) both $g$ and $\bg$ have got  the same sign; whereas for $\kappa^2 A^2-1 <0$ they have opposite sign: there is a change from a Euclidean (Riemannian) metric to a Lorentzian (Pseudo-Riemannian)  one.
\par
Next we will study the expression of the Ricci tensor, its properties when applying the condition \eqref{c}, and  conclude with an illustrative example. 
\par
First, the connection of the metric \eqref{metricg} is
\be \Gamma^\l_{\m\n}=\bar{\Gamma}^\l_{\m\n}+C^\l_{\m\n}\ee
where
\bea &&C^\l_{\m\n}=\frac{\kappa^2}{2}\left[A_\m\bn_\n A^\l-\xi A^\l\bn_\n A_\m+A_\n\bn_\m A^\l-\xi A^\l\bn_\m A_\n-\bn^\l(A_\m A_\n)-\xi\kappa^2A^\l A^\t\bn_\t(A_\m A_\n)\right]\nonumber\\\eea
with $C^\l_{\l\m}=0$ 
The Ricci tensor of the deformed metric  is
\be R_{\m\n}=\bR_{\m\n}+R_{\m\n}^{\text{\tiny{(1)}}}+R_{\m\n}^{\text{\tiny{(2)}}}+R_{\m\n}^{\text{\tiny{(3)}}}\ee
where to be specific,
\bea
R_{\m\n}^{\text{\tiny{(1)}}}&&=\frac{\kappa^2}{2}\Big[FP_{\m\n}-(1+\xi)\left[\bn_\l[A^\l\bn_\m A_\n]+\bn_\l[A^\l\bn_\n A_\m]-\bn_\l A_\m \bn^\l A_\n+\bn_\m A^\l\bn_\n A_\l\right]\Big]\nonumber\\
R_{\m\n}^{\text{\tiny{(2)}}}&&=\frac{\kappa^4}{2}\Big[A_\m A_\n\left(\bn_\a A_\b\bn^\a A^\b-\bn_\a A_\b\bn^\b A^\a\right)-\xi A^\a\bn_\a(A_\m A_\n)\bn_\b A^\b-\nonumber\\
&&-\xi A^\a A^\b\left(A_\m\bn_\a\bn_\b A_\n+A_\n\bn_\a\bn_\b A_\m+\bn_\a A_\m\bn_\b A_\n\right)\Big]\nonumber\\
R_{\m\n}^{\text{\tiny{(3)}}}&&=\xi\frac{\kappa^6}{2}A_\m A_\n A^\a A^\b\bn_\a A^\l\bn_\b A_\l\eea

It is easy to see that now, as opposed to what happens in the null case, FP is not enough to map Ricci flat spaces into Ricci flat spaces. 
\be\label{teorema1}
\{ R_{\m\n}^{\text{\tiny{(1)}}}=0\} \nRightarrow \{R_{\m\n}^{\text{\tiny{(2)}}}+R_{\m\n}^{\text{\tiny{(3)}}}=0\}
\ee
then in general
\be R_{\m\n}\neq\bR_{\m\n}\ee

We shall present an explicit example of this fact in a moment. Start  with the Euclidean metric
\be \text{d}s^2=-2\text{d}t\text{d}z- 2\text{d}x\text{d}y\ee
with $\bR_{\m\n}=0$. Add the deformation
\be h_{\m\n}=\begin{pmatrix}
	\frac{(\chi-2C_1h_2[t])^2}{4C^2_3}& \frac{C_1(\chi-2C_1h_2[t])}{2C_3}& \frac{h_2[t](\chi-2C_1h_2[t])}{2C_3}&\frac{1}{2}\left(\chi-2C_1h_2[t]\right)\\
	\frac{C_1(\chi-2C_1h_2[t])}{2C_3}& C_1^2& C_1h_2[t]&C_1C_3\\
	\frac{h_2[t](\chi-2C_1h_2[t])}{2C_3}& C_1h_2[t]& h_2^2[t]&h_2[t]C_3\\
	\frac{1}{2}\left(\chi-C_1h_2[t]\right)&C_1C_3&h_2[t]C_3&C_3^2
\end{pmatrix} \ee
where the function $h_2[t]$ has the property $h_2^{\prime\prime}[t]=0$ and $\chi\equiv 1+\frac{1}{\xi}$.
This deformation $h_{\m\n}$ obeys  the Fierz-Pauli equation
\be FP_{\m\n}=0\ee
The total metric, with signature $g=1-\chi$
\bea&&\text{d}s^2=-2\text{d}t\text{d}z- 2\text{d}x\text{d}y+\left[\frac{\chi-2C_1h_2[t]}{2C_3}\text{d}t+C_1\text{d}x+h_2[t]\text{d}y+C_3\text{d}z\right]^2\eea
is not Ricci flat , however
\be R_{\m\n}=\frac{(h_2^{\prime}[t])^2}{2-2\chi}\begin{pmatrix}
	C_1^2 & 0& -C_1C_3&0\\
	0 & 0 & 0&0\\
	-C_1C_3 & 0& C_3^2&0\\
	0&0&0&0
\end{pmatrix}\ee 

A natural question to ask at this moment is the following. What are the conditions
that the deformation must obey (we already know that it is not true in general) in order
to keep Ricci flatness? In particular this would hold true whenever
\be R_{\m\n}=\bR_{\m\n}\ee
We will answer this question in the next section.

\section{The general deformation that obeys  $R_{\m\n}=\bR_{\m\n}$}
In this section we want to obtain the necessary conditions that a deformation must fulfill
in order for it to map vacuum solutions into vacuum solutions. Let us start with the necessary but not sufficient condition that the scalar curvature be invariant $R=\bR$. In fact,
\bea &&R=\bR-\kappa^2\xi\left[\bn_\a A^\a\bn_\b A^\b+A^\a\bn_\a\bn_\b A^\b+A^\a\bn_\b\bn_\a A^\b-A^\a\bn_\b\bn^\b A^\a\right]+\nonumber\\
&&+\frac{\kappa^2}{2}(-1+\xi)\left[\bn_\a A_\b\bn^\a A^\b-\bn_\a A_\b\bn^\b A^\a\right]-\xi\frac{\kappa^4}{2}A^\m A^\n\bn_\m A_\l \bn_\n A^\l\eea
Assume that the perturbation satisfies the Fierz-Pauli equation
\be FP_{\m\n}\equiv \frac{1}{2}\left[\bn_\l\bn_\m (A_\n A^\l)+\bn_\l\bn_\n (A_\m A^\l)-\bar{\Box}(A_\m A_\n)\right]=0\ee
multiplying with $A_\m A_\n$
\be\label{FPH} A_\a\bn_\b\bn^\b A^\a-A_\a\bn_\b\bn^\a A^\b=\frac{\xi}{1+\xi}\kappa^2A^\m A^\n\bn_\m A_\l \bn_\n A^\l\ee
taking the trace, that is, multiplying  with $\bg^{\m\n}$
\be \bn_\a A^\a\bn_\b A^\b+A^\a\bn_\a\bn_\b A^\b+A^\a\bn_\b\bn_\a A^\b-A_\a\bn_\b\bn^\b A^\a+\bn_\a A_\b\bn^\b A^\a-\bn_\a A_\b\bn^\a A^\b=0\ee
plugging inside the scalar of curvature both results we learn that
\be R=\bR+\frac{\kappa^2}{2}(1+\xi)\bn_\a\left[A^\b\bn_\b A^\a\right]\ee

This means that in order to have $R=\bR$, we have three options, either  $\xi=-1$, which  corresponds with the null Kerr-Schild or else
\be 
\dot{A}_\b\equiv A^\a \bn_\a A_\b=0\ee
or finally the third possibility, 
\be \dot{ A}_\b=\phi A_\b\hspace{1cm}\text{with}\hspace{1cm}\bn_\a A^\a=0\ee
Let us study the new conditions in turn.
\subsection{The first condition $ \dot{A}_\b=0$ (vanishing rotation).}
To begin with,  automatically $R_{\m\n}^{\text{\tiny{(3)}}}=0$, 
and our previous analysis of the FP equation 
imply that
\be \bn_\b A_\a\bn^\b A^\a-\bn_\b A_\a\bn^\a A^\b=0
\ee
This means that either
\be
\bn_\b A_\a=0
\ee
(in which case the deformed spacetime is automatically Ricci flat) or else
\be
\bn_\m A_\n=\bn_\n A_\m\ee
This means that the flow induced by $A_\m$ is {\em irrotational}, and in fact {\em geodesic} (because $A^2$ is constant). Then
\be
R_{\m\n}^{\text{\tiny{(1)}}}=\frac{\kappa^2}{2}\Big[FP_{\m\n}-(1+\xi)\left[\bn_\l[A^\l\bn_\m A_\n]+\bn_\l[A^\l\bn_\n A_\m]\right]\Big]=
-\xi\frac{\kappa^2}{2}FP_{\m\n}=0\ee
as well as
\be
R_{\m\n}^{\text{\tiny{(2)}}}=-\xi\frac{\kappa^4}{2} A^\a A^\b\left(A_\m\bn_\a\bn_\b A_\n+A_\n\bn_\a\bn_\b A_\m\right)=-\xi\frac{\kappa^4}{2} A^\a A^\b\bn_\a\bn_\b(A_\m A_\n)=0\ee

to conclude
\be FP_{\m\n}=0 \hspace{0.2cm}\text{with}\hspace{0.2cm}\bn_\a A_\b=\bn_\b A_\a \Rightarrow R_{\m\n}=\bR_{\m\n}\ee

this means that whenever the deformation is irrotational, it maintains Ricci flatness.
\subsection{The second alternative $ \dot{A}_\b=\phi A_\b$ plus $\bn_\m A^\m=0$}
This is even simpler. Given the fact that $\ona_\m A^2=0$ the acceleration is orthogonal to the deformation
\be
\dot{A}_\m A^\m=0
\ee
Then $\phi=0$ for any non-null deformation, and we are back on the previous case.

\section{Some examples.}
Let us now consider some examples.

\subsection{Schwarzschild }

 Consider first the background $\bg_{\m\n}$
\be \text{d}s^2=\left(1-\frac{r_s}{r}\right)\text{d}t^2-\frac{1}{\left(1-\frac{r_s}{r}\right)} \text{d}r^2-r^2\left(\text{d}\theta^2+\sin^2{\theta}\text{d}\phi^2\right)\ee

It is not difficult to show that there is a unique deformation of the form
\be \kappa A_\m=\left(f[r],g[r],0,0\right)\ee
that obeys the conditions  of our theorem namely
\be \kappa A_\m=\left(\sqrt{-1-\frac{1}{\xi}},-\frac{1}{1-\frac{r_s}{r}}\sqrt{-\left(1+\frac{1}{\xi}\right)\frac{r_s}{r}},0,0\right)\ee

The total metric reads
\be g_{\m\n}=\begin{pmatrix}
	-\frac{1}{\xi}-\frac{r_s}{r}& \left(1+\frac{1}{\xi}\right)^2\frac{\sqrt{rr_s}}{r-r_s}&0&0\\
	\left(1+\frac{1}{\xi}\right)^2\frac{\sqrt{rr_s}}{r-r_s}& -\frac{r\left(r+\frac{r_s}{\xi}\right)}{(r-r_s)^2}& 0&0\\
	0& 0& -r^2&0\\
	0&0&0&-r^2\sin^2\theta
	\epm
	\ee
Its curvature invariants are
\bea
&R=0\nonumber\\
&R_{\m\n}R^{\m\n}=0\nonumber\\
&R_{\m\n\r\s}R^{\m\n\r\s}=\xi^2\frac{12r_s^2}{r^6}
\eea
The asymptotic limit of $r\rightarrow\infty$, we have $g_{\m\n}=\left(-\frac{1}{\xi},-1,-1,-1\right)$. Note that the value $\xi=-1$  corresponds to lightlike vectors already considered in \cite{Alvarez}.
\subsection{pp-waves }
Consider instead   pp-waves as our background 
\be \text{d}s^2=2\text{d}u\text{d}v-H[u]\text{d}u^2-\text{d}x^2-\text{d}y^2\ee
This metric is known to be Ricci flat.

Let us consider the geodesic deformation
\be \kappa A_\m=\left(0,0,\frac{\sqrt{2}f[u]}{\sqrt{f^2[u]+g^2[u]}},\frac{\sqrt{2}g[u]}{\sqrt{f^2[u]+g^2[u]}}\right)\ee
in this case
\be \kappa^2\bg^{\a\b}A_\a A_\b=-2\ee
this deformation  satisfies the Fierz-Pauli equation , but the total metric
\be \text{d}s^2=2\text{d}u\text{d}v-H[u]\text{d}u^2-\text{d}x^2-\text{d}y^2+\left(\frac{\sqrt{2}f[u]}{\sqrt{f^2[u]+g^2[u]}}\text{d}x+\frac{\sqrt{2}g[u]}{\sqrt{f^2[u]+g^2[u]}}\text{d}y\right)^2\ee
is not Ricci flat.  The  non null component being
\be R_{uu}=\frac{2}{(f^2[u]+g^2[u])^2}\left(f^{\prime}[u]g[u]-f[u]g^{\prime}[u]\right)\ee
 This does not contradict our theorem, because the deformation is only irrotational whenever
\be f^{\prime}[u]g[u]-f[u]g^{\prime}[u]=0\ee
that is, the functions $f$ and $g$ are proportional, in which case
 the total metric is indeed Ricci flat.


\section{Conclusions}
We have generalized the usual null  Kerr-Schild \cite{Kerr} gauge
\be
g_{\m\n}=\bg_{\m\n}+\kappa^2 A_\m A_\n
\ee
with $A^2=0$, to the non null case, allowing any value for $A^2$. This needs to assume that the inverse metric is given by
\be
g^{\m\n}=\bg^{\m\n}+\xi \kappa^2 A^\m A^\n
\ee
where 
\be
A^\l A_\l=-\frac{1}{\kappa^2 }\left(1+\frac{1}{\xi}\right)
\ee
The curvature tensor is now given by a {\em finite} sum of terms, but in contraposition to the null case, however,  the said Fierz-Pauli equation is not enough to map Ricci flat spaces amongst themselves. 
\par
We have been able to prove a theorem however, that states that when the deformation is irrotational (which implies that it is also geodesic, because $A^2$ is constant)  with respect to the background metric, that is when
\be
\ona_\m A_\n =\ona_\n A_\m
\ee
then the Ricci tensor is fully determined by its linear piece, so that if the deformation obeys the  the Fierz-Pauli equation
then the Ricci tensor is invariant. This obviously means that whenever  the background spacetime is Ricci flat, so is the deformed one.
\be
\bR_{\m\n}=R_{\m\n}
\ee
Given a Ricci flat  background {\em seed} metric any irrotational  deformation will lead to a different spacetime, also Ricci flat. 
An amusing thing is the following. If we start with a constant curvature background metric, {\em id est}
\be
\overline{R}_{\m\n}= {1\over 4}\,\oR \og_{\m\n}
\ee
then the deformed metric obeys
\be
R_{\m\n}=\oR_{\m\n}={1\over 4}\,\oR \og_{\m\n}={1\over 4}\oR\,\left(g_{\m\n}-\kappa^2 A_\m A_\n\right)
\ee
which corresponds to an (albeit strange) perfect fluid as a source.
\par

It is likely that some of these ideas are also useful in the double copy \cite{Gurses} framework.
We are planning  a systematic study of this topic  in  future work.

\section{Acknowledgements}
We  acknowledge partial financial support by the
Spanish MINECO through the Centro de excelencia Severo Ochoa Program  under Grant CEX2020-001007-S  funded by MCIN/AEI/10.13039/501100011033.
We also acknowledge partial financial support by the Spanish Research Agency (Agencia Estatal de Investigaci\'on) through the grant PID2022-137127NB-I00 funded by MCIN/AEI/10.13039/501100011033/ FEDER, UE.
All authors acknowledge the European Union's Horizon 2020 research and innovation programme under the Marie Sklodowska-Curie grant agreement No 860881-HIDDeN and also by Grant PID2019-108892RB-I00 funded by MCIN/AEI/ 10.13039/501100011033 and by ``ERDF A way of making Europe''.


  
\end{document}